\documentstyle[charm2000]{article}
\def\Journal#1#2#3#4{{#1} {\bf #2}, #3 (#4)}

\def\PLB{{\em Phys. Lett.}  B}
\def\PRL{\em Phys. Rev. Lett.}
\def\PRD{{\em Phys. Rev.} D}

\def\be{\begin{equation}}
\def\ee{\end{equation}}
\def\bea{\begin{eqnarray}}
\def\eea{\end{eqnarray}}

\begin{document}

\title{
\begin{flushright}
PRINCETON/HEP/97-5 \\
hep-ph/9706477 \\
\end{flushright}
\large\bf An Overview of $D^0\bar{D}^0$ Mixing Search
Techniques~\thanks{ Presented at Symposium on Flavor Changing
Neutral Currents: Present and Future Studies,
Santa Monica, CA. Feb. 19-21, 1997.}
}
\author{\large{Tiehui (Ted) Liu}                
     \\ \\ {\sl Department of Physics, Princeton University, 
Princeton, NJ 08544}    
         }  


\maketitle

\begin{abstract}
The search for $D^0\bar{D}^0$
mixing may carry a large discovery potential for new 
physics since the $D^0\bar{D}^0$ 
mixing rate is expected to be small in the Standard Model. 
This paper gives a brief review of the experimental techniques which
can be used to search for $D^0\bar{D}^0$ mixing.
\end{abstract}
  
\section{Introduction}

Much of the enthusiasm for searching for $D^0\bar{D}^0$ 
mixing stems from the belief that the search carries a large discovery 
potential for New Physics, since the mixing rate 
${\rm R_{mixing}} \equiv {\cal B}(D^0 \to \bar{D}^0 \to \bar{f})$/
${\cal B}(D^0 \to f)$ is expected to be small in the Standard Model
(range from $10^{-7}$ to $10^{-3}$, see references 
in elsewhere~\cite{Taucharm,pakvasa,Golowich}).
One can characterize $D^0\bar{D}^0$ mixing in terms of two
dimensionless variables: 
$ x={\delta m / \gamma_+}$ and $y={\gamma_- / \gamma_+}$, where
the quantities $\gamma_\pm$ and $ \delta m$ are defined by 
$\gamma_\pm = {(\gamma_1\pm \gamma_2) / 2}$ and $ \delta m = m_2 - m_1$
with $m_i,\gamma_i$ $(i=1,2)$ being the masses and decay rates of 
the two CP (even and odd) eigenstates. 
Assuming a small mixing, namely,
$\delta m, \gamma_- \ll \gamma_+$ or $x,y\ll 1$, we have
${\rm R}_{\rm mixing}= (x^2 + y^2)/2 $.
An overview of the current experimental status and
future prospects can be found elsewhere~\cite{Taucharm}.
This paper only gives a brief review of the experimental techniques which
can be used to search for $D^0\bar{D}^0$ mixing.


\section{The Techniques} 

The techniques which can be used to search for mixing can be roughly 
divided into two classes: hadronic and semi-leptonic. Each method has 
advantages and limitations, which are described below. 

\subsection{Hadronic Method}
The hadronic method is to search for the $D^0$ decays 
$D^0\to K^+\pi^-(X)$~\cite{footnote}.
These decays can occur either through 
$D^0\bar{D}^0$ mixing followed by Cabibbo favored
decay $D^0 \to \bar{D}^0  \to K^+\pi^-$(X), or 
through DCSD (Doubly Cabibbo Suppressed Decay)
$D^0\to K^+\pi^-(X)$. 
This means that the major complication for this 
method is the need to distinguish between DCSD and mixing. 
The hadronic method can therefore be classified according to how 
DCSD and mixing are distinguished. In principle, there are 
at least three different ways to distinguish between DCSD and
mixing candidates experimentally: 
(A) use the difference in the decay time-dependence;
(B) use the possible difference in the resonant substructure 
    between DCSD and mixing events
    in $D^0 \to K^+\pi^-\pi^0$,$ K^+\pi^-\pi^+\pi^-$, etc. modes; 
(C) use the quantum statistics of the production 
and the decay processes; (D) there is also a possibility 
of using the CP eigen states, such as $D^0 \to K^+K^-,\pi^+\pi^-$ 
(Singly Cabibbo Suppressed Decays -- SCSD), to study mixing
caused by the decay rate difference:
$y={\gamma_- / \gamma_+}=(\gamma_2 - \gamma_1) /(\gamma_1 + \gamma_2)$.
Note that DCSD decays are interesting in their own right,
since they can be used to extract CKM angle $\gamma$ in 
$B^- \to D^0(\bar{D^0})K^-$ modes~\cite{atwood}.


\subsubsection{Method A --use the 
difference in the time-dependence of the decay}

This method is to measure the decay time of the $D^0 \to K^+\pi^-$ decay.
Here the $D^0$ tagging is usually done by using 
the decay chain $D^{*+} \to D^0{\pi}_{\rm s}^+$ followed by 
$D^0 \to K^+ \pi^-$. 
A pure $D^0$ state generated at $t=0$  decays to the $K^+\pi^-$ state either 
by $D^0\bar{D}^0$ mixing or by DCSD, and the 
two amplitudes may interfere. 

Following the notation in Ref.1,
and define 
$a(f) = Amp(D^0 \to f)$, $\bar{a}(f) = Amp(\bar{D^0} \to f)$
with $ \rho (f)= a(f)/\bar{a}(f)$;
$a(\bar{f}) = Amp(D^0 \to \bar{f})$, 
$\bar{a}(\bar{f}) = Amp(\bar{D^0} \to \bar{f})$
with $\bar{\rho}(\bar{f})= \bar{a}(\bar{f})/a(\bar{f})$,
and $\eta \equiv \frac{p}{q}\rho(f)$
($\bar{\eta} \equiv \frac{q}{p}\bar{\rho}(\bar{f})$ ),
for $\delta mt,\delta{\gamma}t << 1$ and small $|\eta|$ (
in the case of $f=K^+\pi^-$), we have~\cite{Taucharm}
\begin{equation}
\label{mixingrate_kpi}
I(~|D^0_{\rm phys}(t)> \to f~) = |\bar{a}(f)|^2~|\frac{q}{p}|^2~|~\frac{1}{2}(i\delta m + 
\frac{1}{2}\delta \gamma )~t + \eta~|^2~ e^{-\gamma_+t}.
\end{equation}

\begin{equation}
\label{mixingrate_kpibar}
I(~|\bar{D^0}_{\rm phys}(t)> \to \bar{f}~) = 
|a(\bar{f})|^2~|\frac{p}{q}|^2~|~\frac{1}{2}(i\delta m + 
\frac{1}{2}\delta \gamma )~t + \bar{\eta}~|^2~ e^{-\gamma_+t}. 
\end{equation}

Note the difference between Equation~\ref{mixingrate_kpi} and~\ref{mixingrate_kpibar}
is the indication of CP violation. 
Next let us write the equations in a somewhat different form,
which is more convenient for discussion here. 
Note that $
|~\frac{1}{2}(i\delta m + \frac{1}{2}\delta \gamma )~t + \eta~|^2 $
can be written in the form: 
$|~\frac{1}{2}(ix+y)~t + \eta~|^2 $ 
where now the time t is measured in unit of average $D^0$ lifetime ($1/\gamma_+$).
Recall that we have ${\rm R}_{\rm mixing} = (x^2 + y^2)/2$ and
define ${\rm R}_{\rm DCSD}=|\eta|^2$, the term above
can be written in the form $|~\sqrt{{\rm R}_{\rm mixing}/2}\:\; t +
\sqrt{{\rm R}_{\rm DCSD}}\:\; e^{i\phi}~|^2$
where $\phi=Arg(ix+y)-Arg(\eta) $ is an unknown phase between the
two amplitudes.

Equation~\ref{mixingrate_kpi} and~\ref{mixingrate_kpibar}
now become:
\begin{eqnarray}
\label{time_dep_kpi_cp}
\lefteqn{I(|D^0_{\rm phys}(t)> \to f) = 
|\bar{a}(f)|^2~|\frac{q}{p}|^2~\times} \nonumber \\ 
 & & {\left[{\rm R}_{\rm DCSD} + 
\sqrt{2{\rm R}_{\rm mixing}{\rm R}_{\rm DCSD}}~t~\cos\phi + 
\frac{1}{2}{\rm R}_{\rm mixing}\:\; t^2~\right]~e^{-t}. }
\end{eqnarray}
and
\begin{eqnarray}
\label{time_dep_kpi_cp_bar}
\lefteqn{I(|\bar{D^0}_{\rm phys}(t)> \to \bar{f}) = 
|a(\bar{f})|^2~|\frac{p}{q}|^2~\times} \nonumber \\
 & & {\left[\overline{{\rm R}}_{\rm DCSD} + 
\sqrt{2{\rm R}_{\rm mixing}\overline{{\rm R}}_{\rm DCSD}}~t~\cos\bar{\phi} + 
\frac{1}{2}{\rm R}_{\rm mixing}\:\; t^2~\right]~e^{-t}. }
\end{eqnarray}
where $\overline{{\rm R}}_{\rm DCSD}=|\bar{\eta}|^2$ and 
$\bar{\phi}=Arg(ix+y)-Arg(\bar{\eta})$. Note $\bar{\phi}$ is
different from $\phi$ because $Arg(\bar{\eta}) = Arg(q/p)
+ Arg(\bar{\rho}(\bar{f}))$ where 
$Arg(q/p)=-Arg(p/q)$
(and one can define $q/p=e^{-2i\phi_{M}}$ in
the usual way).
For convenience, let us ignore the CP violation in the
decay amplitude. Thus the difference between the interference
terms (that is, the interference phase $\phi$ and $\bar{\phi}$)
in Equation~\ref{time_dep_kpi_cp} and~\ref{time_dep_kpi_cp_bar}
would be the indication of CP violation. 

Assuming CP conservation, Equation~\ref{time_dep_kpi_cp} 
and~\ref{time_dep_kpi_cp_bar} simply become~\cite{Charm2000}:

\begin{equation}
\label{time_dep_kpi}
{\rm I}(|D^0_{\rm phys}(t) \to K^+\pi^-)(t) \propto ({\rm R}_{\rm DCSD} 
+\sqrt{2{\rm R}_{\rm mixing}{\rm R}_{\rm DCSD}}\;\: t\: cos\phi
+\frac{1}{2}\;{\rm R}_{\rm mixing}t^2)e^{-t}.
\end{equation}

The signature of mixing is therefore a deviation from a perfect
exponential time distribution with the slope of $\gamma^+$.
A small mixing signature could show up in the
interference term at lower decay times.
The importance of the interference term has been discussed
in detail elsewhere~\cite{Charm2000,Taucharm}. 
The importance of possible large CP violation
effect (due to New Physics) has been discussed 
by Wolfenstein~\cite{CPinmixing1} and others~\cite{CPinmixing2}.
It has also recently been argued by Browder and Pakvasa~\cite{Tom}
that it may be possible to calculate the phase
$Arg(\rho(f))$ due to final state interaction.

It is interesting to take a look at the time integrated effect. Let us define
\begin{equation}
\label{time_int_kpi_define}
{\rm R}=\frac{\int_{0}^{\infty} I(|D^0_{\rm phys}(t)> \to f)\,dt}
{\int_{0}^{\infty} I(|\bar{D^0}_{\rm phys}(t)> \to f)\,dt}.
\end{equation}
and we have~\cite{Taucharm}
\begin{equation}
\label{time_int_kpi}
{\rm R}= {\rm R}_{\rm DCSD} 
+\sqrt{2{\rm R}_{\rm mixing}{\rm R}_{\rm DCSD}}\; \cos\phi
+ {\rm R}_{\rm mixing}.
\end{equation}

In the special case when $|\cos\phi|=1$, Equation~\ref{time_int_kpi}
can be written in the form
\begin{equation}
\label{time_int_kpi_sp}
(~{\rm R}_{\rm DCSD}-{\rm R}~)^2 
+ (~{\rm R}_{\rm mixing}-{\rm R}~)^2 = {\rm R}^2.
\end{equation}

\subsubsection{Method B --use the 
difference in the resonant substructure between
DCSD and mixing events in multi-body decays}

The idea~\cite{Charm2000} is to use the wrong sign 
decay $D^{*+} \to D^0{\pi}_{\rm s}^+$ followed by 
$D^0 \to K^+ \pi^-\pi^0$, $K^+\pi^-\pi^+\pi^-$, etc., and use the 
possible differences of the resonant substructure 
between mixing and DCSD to study mixing.
In principle, one can use the difference between the 
resonant substructure for DCSD and mixing events to 
distinguish mixing from DCSD. For instance, 
combined with method A, one can perform a multi-dimensional
fit to the data by using the information on 
$\Delta M$, $M(D^0)$, proper decay time $t$ and 
the yield density on Dalitz plot $n_{w}(p,t)$. The extra information on the
resonant substructure will, in principle, put a better 
constraint on mixing.
Of course, one needs large amount of clean data to
do this in the future. 
Because of this, for current experiments
this method is more likely to be a complication
rather than a better method.  
In general, we cannot treat multi-body decays 
exactly the same way as $D^0 \to K^+ \pi^-$ when
the resonant substructure is ignored.
One can find detail discussions about this  
elsewhere~\cite{Taucharm}.

\subsubsection{Method C ---use quantum statistics of the production 
and decay processes}
This method is to search for dual identical two-body hadronic 
decays in $e^+e^- \to \Psi'' \to D^0\bar{D}^0$, such as 
$(K^-\pi^+)(K^-\pi^+)$, as was first suggested by Yamamoto
in his Ph.D thesis~\cite{Yamamoto}.
The idea is that when $D^0\bar{D}^0$ pairs are generated in a state
of odd orbital angular momentum (such as $\Psi''$), the DCSD
contribution to identical two-body pseudo-scalar-vector ($D \to PV$)
and pseudo-scalar-pseudo-scalar ($D \to PP$) hadronic decays 
(such as $(K^-\pi^+)(K^-\pi^+)$) cancels out, leaving only the contribution
of mixing~\cite{Yamamoto,Bigi} in the wrong sign versus right
sign ratio:
\begin{equation}
\label{nodcsd}
{\rm R}=\frac{{\rm N}(K^-\pi^+,K^-\pi^+)+{\rm N}(K^+\pi^-,K^+\pi^-)
}{{\rm N}(K^-\pi^+,K^+\pi^-)+{\rm N}(K^+\pi^-,K^-\pi^+) } 
=(x^2 + y^2)/2 = {\rm R}_{\rm mixing}.
\end{equation}
One common misunderstanding
about this method is that DCSD is forbidden in this case. This
is not true, DCSD is allowed with mixing and contributes to both
wrong sign and right sign (but the contribution cancels out in the 
ratio). One can find the essence of Yamamoto's original calculation for the
$(K^-\pi^+)(K^-\pi^+)$ case elsewhere~\cite{Taucharm}. 
Detailed calculations for this kind of method can be found in a recent 
paper~\cite{xing}.

\subsubsection{Method D -- use CP eigen final states
(such as $D^0 \to K^+K^-,\pi^+\pi^-$ )
to measure the decay rate difference $y$}

This is because (assuming CP conservation)
those decays occur only through the CP even eigenstate,
which means the decay time distribution is a perfect 
exponential with the slope of $\gamma_1$.
Therefore, one can use those modes to measure $\gamma_1$.
The mixing signature is not a deviation from
a perfect exponential (assuming CP conservation), 
but rather a deviation of the
slope from $(\gamma_1 + \gamma_2)/2$. Since 
${\gamma_+}= (\gamma_1 + \gamma_2)/2$ can be measured by 
using the $D^0 \to K^-\pi^+$ decay time distribution, one can then
derive $y={\gamma_- / \gamma_+}=(\gamma_2 - \gamma_1) /(\gamma_1 + \gamma_2)$.
Observation of a non-zero $y$ would demonstrate mixing caused
by the decay rate difference
(${\rm R}_{\rm mixing}= (x^2 + y^2)/2 $). 
Note that there is no need to tag the $D^0$, since we only need to determine
the slope.

\subsection{Semi-Leptonic Method}

The semi-leptonic method is to search for 
$D^0\to \bar{D^0}\to X l^- \nu$ decays, 
where there is no DCSD involved. However, it usually (not always)
suffers from a large background due to the missing neutrino.
In addition, the need to understand the large background often 
introduces model dependence. New ideas are needed in order to improve the
sensitivity significantly. One can find more discussions
elsewhere~\cite{Taucharm,Morrison}.

\section{Comments On Experimental Issues}

Generally speaking, vertexing and tagging are
the two critical elements to the study
of mixing and CP violation. For $B_{d}\bar{B_{d}}$ and $B_{s}\bar{B_{s}}$,
tagging has always been one of the major issues and has been
proven to be difficult. 
In $D^0\bar{D}^0$ case, 
tagging seems to be easier as one has the clear advantage of
using the decay chain $D^{*+} \to D^0{\pi}_{\rm s}^+$ (see below),
however, the much smaller $D^0\bar{D}^0$ mixing rate requires much cleaner
tagging.  
 
As mentioned before, the $D^0$ tagging is usually done by using 
the decay chain $D^{*+} \to D^0{\pi}_{\rm s}^+$ followed by 
$D^0 \to K^+ \pi^-$. The ${\pi}_{\rm s}^+$ from 
$D^{*+}$ has a soft momentum spectrum and is referred to as ``the slow pion''.
The charge of the slow pion is correlated with the charm quantum number
of the $D^0$ meson and thus can be used to tag whether a $D^0$ or 
$\bar{D}^0$ meson was produced.
The idea is to search for wrong sign $D^{*+}$ decays, where the
slow pion has the same charge as the kaon arising from the
$D^0$ decay. The right sign signal $D^{*+} \to D^0{\pi}_{\rm s}^+$ 
followed by $D^0 \to K^- \pi^+$ can be used to provide
a model-independent normalization for the mixing measurement. 
The small $Q$ value of the $D^{*+}$ decay
results in a very good mass resolution in the mass difference 
$\Delta M \equiv M(D^{*+}) - M(D^0) - M({\pi}_{\rm s}^+)$ and allows 
a $D^{*+}$ signal to obtained with very low background. 
In general, mixing search requires at least: 
(1) excellent vertexing capabilities, at least good enough
to see the interference structure; 
(2) low background around the primary vertex.

The vertexing capabilities at $e^+e^-$ experiments
(${\cal L}/\sigma \sim 3$) for CLEO III and asymmetric B factories
at SLAC and KEK may be sufficient for a mixing search.
The extra path length due to the Lorentz boost,
together with the use of silicon detectors for high
resolution position measurements, have given the fixed 
target experiments an advantage in vertex resolution
(typically ${\cal L}/\sigma \sim 8-10$)
over $e^+e^-$ experiments. 

The background level around the primary vertex
could be an important issue. 
Low background around primary vertex 
means that one does not suffer much from random slow pion background. 
Define $B$ as the number of wrong sign background events, $S$ the number of
signal events and $Q$ the $D^* - D$ mass difference,
one has~\cite{Morrison}:

\begin{equation}
\label{bovers}
\frac{B}{S} \sim (\frac{1}{S}\frac{dB}{dQ})(2\sigma_{Q})
\end{equation}

where $\sigma_{Q}$ is the mass difference resolution.
Note the background density, $\frac{1}{S}\frac{dB}{dQ}$,
is a characteristic of the fragmentation process.
At CLEO II, the background density is about 0.001 per $MeV$
and $\sigma_{Q} \sim 0.7 MeV$ which means that
$B/S \sim 10^{-3}$. For CLEO III and $BF$,
$\sigma_{Q}$ can be reduced down to $0.3 MeV$ with
the silicon vertex detector. 
The low background around the primary vertex at 
$e^+e^-$ experiments is a certain advantage.
One disadvantage at fixed target experiments is 
the higher background around the primary vertex 
(higher $B/S$).

While vertexing
capabilities will likely be improved in the future,
the background density
seems difficult 
to improve. However, there are some
ideas~\cite{Charm2000} 
to improve the background density
for the future experiments. For example,
for asymmetric B factory or Z factory, fixed target experiments
(such as HEAR-B) and especially hadron machines (TEVATRON, LHC),
it maybe possible to use $\bar{B^0} \to D^{*+} l^-\nu $, where the
primary ($D^{*+}$ decay) vertex can be determined by the $l^-$ together with 
the slow pion coming from the $D^{*+}$. In this case, the background level
around the primary vertex is intrinsically low since the
$D^*$ (or $B$) decay vertex is no longer the event vertex.
Here the background density, $\frac{1}{S}\frac{dB}{dQ}$,
is no longer a characteristic of the fragmentation process.
In fact, CDF~\cite{CDF} has recently obtained rather clean right sign signal
for $\bar{B^0} \to D^{*+} l^-\nu$ followed by the 
decay chain $D^{*+} \to D^0{\pi}_{\rm s}^+$ and 
$D^0 \to K^- \pi^+$ ($\sim 1000 $ signal events from
$110 {\rm pb}^{-1}$ data).

It maybe also possible to use hadronic $B$ decays (assuming
one can trigger on these events),
such as $\bar{B^0} \to D^{*+} \pi^-$ ($BR \sim 0.3\%$),
$\bar{B^0} \to D^{*+}\pi^-\pi^0$ ($BR \sim 1.5\%$),
$\bar{B^0} \to D^{*+}\pi^-\pi^+\pi^-$ ($BR \sim 1.2\%$) and
$\bar{B^0} \to D^{*+}\pi^-\pi^+\pi^-\pi^0$ ($BR \sim 3.4\%$). 
In addition, charged $B$ decays such as 
$B^- \to D^{*+}\pi^-\pi^-$ ($BR \sim 0.2\%$),
$B^- \to D^{*+}\pi^-\pi^-\pi^0$ ($BR \sim 1.5\%$) can
be also used. There should be enough tracks which can
be used to determine the $B$ (or $D^{*+}$) decay vertex.
Here not only is the background level
around the $D^{*+}$ primary vertex intrinsically low,
but also the backgrounds can be further reduced
by the $B$ mass constraint as one can fully reconstruct
the $B$ decays (unlike in the $\bar{B^0} \to D^{*+} l^-\nu$ case).
Note that this is similar to the idea~\cite{c0workshop}
of using $B_{c} \to B_{s}\pi^+ (\rho^+)$ to tag $B_{s}$.
One major difference is that $B_{c}$ production
cross section is unknown and could be very low,
while one expects more than $10^{10}$ $B$ produced
at Tevatron and LHC which can be potentially used for 
$D^0\bar{D}^0$ mixing search.

\section{Summary}

We have briefly discussed some possible techniques which can be used for
mixing searches in the future. They can be roughly
divided into two classes: hadronic and semi-leptonic.  
Each method has advantages and limitations. 
In the case of semi-leptonic method, the advantage
is that it is theoretically clean as there is
no DCSD involved while the limitation is that very 
often it is limited by large background due to the
missing neutrino.
In the case of $D^0 \to K^+\pi^-(X)$, the major
complication is the presence of DCSD. However, 
this very complication can be turned to advantage
since the potentially small mixing signature could show up 
in the interference term (at lower decay times).
The design of future experiments should focus on improving the 
vertexing capabilities and reducing the
background level around the primary vertex, in order to
fully take advantage of having the possible DCSD and mixing interference. 
In addition, we have learned that the very complication due to 
the possible differences between the
resonant substructure in many DCSD and mixing decay modes 
$D^0 \to K^+\pi^-X$ could also, in principle, be turned to advantage by
providing additional information.

In the case of $D^0 \to K^+\pi^-(X)$ and $D^0 \to X^+ l^-$, we are only
measuring ${\rm R}_{\rm mixing}= (x^2 + y^2)/2 $.
Since many extensions of the Standard Model predict large
$ x={\delta m / \gamma_+}$, and we expect New Physics does not
affect the decays in a significant way thus does not
contribute to $y$, it is important to measure
$x$ and $y$ separately. Fortunately, decays 
such as $D^0 \to K^+K^-,\pi^+\pi^-$, can provide us information
on $y$. Observation of a non-zero $y$ would demonstrate mixing caused
by the decay rate difference.
This, together with the information on 
${\rm R}_{\rm mixing}$ obtained from other methods, we can
in effect measure $x$. It is worth to point out here that $x$ and $y$
are expected to be at the same level within the Standard Model, however
we do not know for sure exactly at what level since theoretical
calculations for the long distance contribution 
are still plagued by large uncertainties. 
Therefore, it is very important to 
measure $y$ in order to understand the size of $x$
within the Standard Model, so that when $D^0\bar{D^0}$ mixing
is finally observed experimentally, we will know whether
we are seeing the Standard Model physics or new physics beyond
the Standard Model.

\section*{Acknowledgement}

I would like to express special thanks to
Lincoln Wolfenstein for many useful and stimulating discussions
at this conference and Honolulu CP conference.
This work is supported by Robert H. Dicke Fellowship Foundation at
Princeton University.



\section*{References}
\begin{thebibliography}{99}






\bibitem{Taucharm}
T. Liu, ``An Overview of $D^0\bar{D^0}$ Mixing Search 
Techniques: Current Status and Future Prospects,''
in Proceedings of the Workshop
on the Tau/Charm Factory, page 447-479,
Argonne National Laboratory, Argonne, IL, 
June 21-23, 1995. Princeton/HEP/95-6, HEP-PH/9508415, and references
therein. 



\bibitem{pakvasa}
S. Pakvasa, these proceedings.

\bibitem{Golowich}
E. Golowich, ``Theory: Rare Decays, Mixing, CPV in $D$ Mesons,''
in Proceedings of the B Physics and CP Violation
Conference, University of Hawaii, Honolulu, March 24-27, 1997.

\bibitem{footnote}We discuss $D^0$ decays explicitly in the text,
its charge conjugate decays are also implied 
throughout the text unless otherwise stated.

\bibitem{atwood}
D. Atwood, I. Dunietz, A. Soni,
\Journal{\PRL}{78}{3257}{1997}.

\bibitem{Charm2000}
T. Liu, ``The $D^0\bar{D^0}$ Mixing Search - Current Status and Future
Prospects,'' in Proceedings of the Charm 2000 Workshop, FERMILAB-Conf-94/190,
page 375-394,
Fermilab, June 7-9, 1994. Harvard preprint HUTP-94/E021, HEP-PH 9408330.


\bibitem{CPinmixing1}
L. Wolfenstein, 
\Journal{\PRL}{75}{2460}{1995}.

\bibitem{CPinmixing2}
G. Blaylock, A. Seiden and Y. Nir, 
\Journal{\PLB}{355}{555}{1995}.

\bibitem{Tom}
T. E. Browder and S. Pakvasa, 
\Journal{\PLB}{383}{475}{1996}.

\bibitem{Yamamoto}
H. Yamamoto, Ph.D thesis, CALT-68-1318 (1985).

\bibitem{Bigi}I. Bigi and A. I. Sanda, 
\Journal{\PLB}{171}{320}{1986}.


\bibitem{xing}
Z. Xing,
\Journal{\PRD}{55}{196}{1997}.


\bibitem{Morrison}
R. J. Morrison, 
``Charm 2000 Workshop Summary,'' in
Proceedings of the Charm 2000 Workshop, FERMILAB-Conf-94/190,
page 313-322, Fermilab, June 7-9, 1994.

\bibitem{CDF}
The CDF II Collaboration,
``The CDF II Detector Technical Design Report,''
FERMILAB-Pub-96/390-E, October, 1996.

\bibitem{c0workshop}
T. Liu and S. Pakvasa,
``Rare Decay Searches at C0,'' in
Proceedings of the Workshop on Heavy Quark Physics at C-Zero,
Fermilab, Dec. 4-6, 1996.








\end{thebibliography}
\end{document}